# A Coupled CMOS Oscillator Array for 8ns and 55pJ Inference in Convolutional Neural Networks

D. E. Nikonov[1], P. Kurahashi[2], J. S. Ayers[2], H.-J. Lee[2], Y. Fan[2], and I. A. Young[1]

[1]Components Research, TMG, Intel, Hillsboro, USA, email: dmitri.e.nikonov@intel.com

[2]Advanced Design, TMG, Intel, Hillsboro, USA

Recently significant research effort has been directed to artificial intelligence and machine learning in order to achieve efficient computing of highly specialized tasks that typically aren't performed well by classical computing methods. Among these tasks is image recognition which can be efficiently computed by an algorithm of a convolutional neural network (CoNN) [1] which enables inference, i.e., determination of how close the input pattern is to a class of memorized patterns. Neural networks for inference, often based on analog computing, are under active exploration and are expected to be more efficient than general purpose computing. The basis of CoNN, see Fig. 1, is a convolution (equivalent to a dot product) with kernels [1] (in this case 5x5 pixel Gabor filters). The example image fragment is closer and thus has a larger convolution with 4 highlighted filters. The gray scale of the pixel is mapped to [-1;1]. Therefore the aim here is to replace the sum of products of 25 real numbers with a single analog operation.

One type of neural networks based computing hardware is oscillator neural networks (ONN), consisting of arrays of coupled oscillators [2]. The input and memorized patterns are encoded as phases of the oscillators, phase shift keying (PSK) [3] or alternatively as the frequencies of the oscillators, frequency shift keying (FSK) [4]. The case when oscillators synchronize corresponds to positive inference (recognition) and the case when they do not, corresponds to a lack of recognition. Chips with PSK CMOS oscillators have been fabricated [5], but neural information processing has not yet been demonstrated. Additionally coupled beyond-CMOS oscillators based on metal-insulator transitions [6] or spin-torque have been proposed for speech recognition and fabricated [7]. However synchronizing more than a few oscillators [6,7] or controlling them individually [8] has proven to be a challenge. In this paper a demonstration of a large array of coupled CMOS oscillators performing inference with special focus on image recognition is shown.

The oscillator array is shown in Fig. 2. Each of the 26 ring oscillators (Fig. 3) is controlled by a 5-bit PMOS IDAC circuit. Each IDAC shares a single bias voltage generated on chip by an external reference current. The 25 "pixel by pixel" differences of the fragment and the kernel (with one unused oscillator to maintain layout symmetry) are mapped to the range of IDAC codes from 0 to 20 and therefore results in the distribution of initial frequencies of oscillators [4] with sufficiently linear dependence, Fig. 4.

The oscillators are coupled together via tunable capacitors to the 'averager' node. Its voltage is rectified and then sampled to determine the "degree of match" (DOM) between the fragment and one of the kernels. Oscillators synchronize only if their frequencies are sufficiently close, i.e., within the 'locking range' determined by the value of the coupling capacitance. A fragment with a strong match to a kernel filter will result in a narrower initial frequency distribution, within the locking range. Hence more oscillators will synchronize, which produce a larger and faster growing voltage at the rectifier output. A poor match between a fragment and a kernel filter will result in a wider initial frequency, outside the locking range. Hence fewer oscillators will synchronize, which produces a lower and slower growing rectifier output.

The amount of coupling capacitance is tuned to achieve the optimum correlation between dot products and oscillator DOM outputs. If the capacitance is too large, too many cases will lead to synchronization. If the capacitance is too small, none of the cases will lead to synchronization. Both limiting cases lead to a lack of distinguishability between high and low convolution values. Tunability is realized with banks of programmable capacitors which can be switched to multiple capacitance values of a few fF, depending on the number of stages. It is critical that the coupling between oscillators in the array is limited to the coupling though the averager and any parasitic uncontrolled coupling from the input bias or supply is minimized. Therefore a passive RC filter is used on the IDAC's input bias in order to achieve approximately 40dB of isolation at the frequencies of oscillation. Additionally, the IDAC has sufficient supply decoupling capacitance to provide a similar isolation from supply coupling, as verified by simulation.

During operation, all oscillators power up at the same time with the enable signal $V_{trig}$. For correct synchronization, all oscillators in the array need to start with random and evenly distributed initial phases. The initial condition of the oscillators is controlled by the n-bit "ic" signal where "n" is the number of inverters. The phase is initialized by breaking the connection between each inverter and forcing the initial output of each inverter to the value determined by "ic". When the loop is enabled, the connections between inverters in the loop are restored and the devices used to drive "ic" are put into a high-impedance state. All parameters are loaded on chip via a 1536-bit scan chain.

The circuit was designed and fabricated in a 22nm CMOS process with nominal supply of 1V. Three versions of oscillator arrays, with 3, 5, and 7 stages, were implemented in order to further study the tradeoffs between oscillator speed and synchronization time. An on-chip DFT divider enables viewing of the waveform of a selected oscillator with its frequency divided by 512. Fig. 4 shows the typical operating frequency of the oscillator outputs vs. IDAC code with the coupling capacitance disabled. Four chips from two wafers were tested.

The DOM is determined from the envelope of the ac signal at the averager, using a rectifying circuit, peak detector (Fig. 3). The peak detector is designed to quickly rectify the oscillations on the averager and to drive the resulting signal off chip. This is achieved by cascading two source follower stages, each of which has a step up in drive current in order to deliver the analog DOM voltage waveforms for recording on an oscilloscope, top Fig. 5. The peak detector is enabled by $V_{t\_del}$ and starts integrating the averager voltage 2.2ns after $V_{trig}$ according to circuit simulations. The sampling time of the DOM off-chip is chosen at 6ns after the start of integration to measure the "degree of match". Circuit simulations and measurements performed on chips (see typical results in Fig. 5) confirm that DOM is well correlated with the ideal convolution/ dot product of the fragment and the kernel. The relationship is very linear (with $R^2$ fit goodness of >0.89).

The inference delay is deduced to be ~8ns (including the 2-stage source follower buffer delay). The simulated average power of each oscillator is 0.26mW and the power of the peak detector is less than 100uW during normal operation with an additional 1.6mA of current used if the DFT circuitry is enabled. Thus the energy per inference is 55pJ. The energy and delay of an inference are benchmarked against published neural accelerator chips [9], Fig. 6. This comparison is not quite equitable, since this work pertains to a single block while others are complete chips. Still the CMOS ONN architecture provides a competitive option for inference hardware [10]. The die photograph in Fig. 7 shows the placement of the circuits on the test chip. Each of the three arrays contains the 26 oscillator and IDAC assemblies and a single peak detector, while the bias and DFT circuitry are shared among the arrays. Each individual array and peak detector occupies an area of 0.01252mm$^2$.

A proof of principle neural network inference engine for a CoNN implemented with an ONN based on a coupled CMOS ring oscillator array is demonstrated. To the authors' knowledge this is the first demonstration of a CMOS oscillator array used for calculating inference for image recognition.

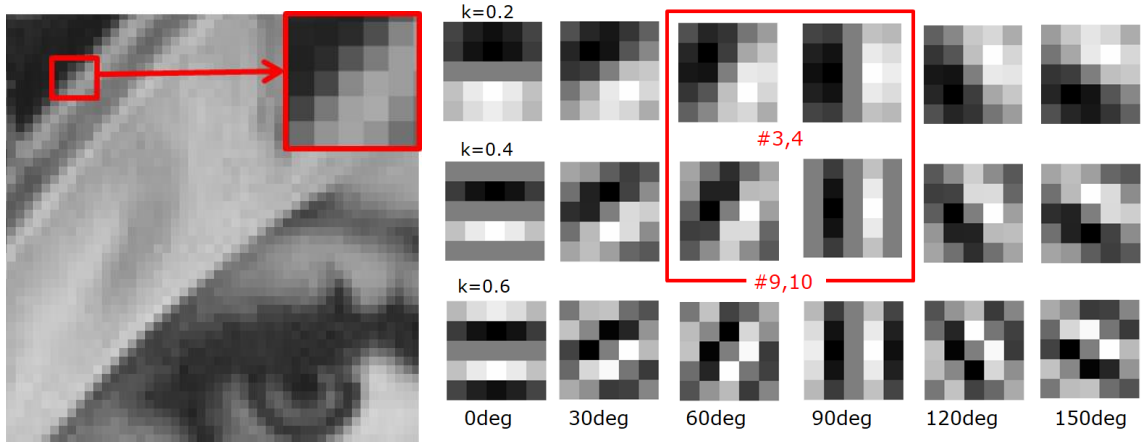

Fig. 1: A fragment of an image and a set of Gabor filters. The direction of the wave-like variation (degrees) and the parameter of the inverse wavelength, k, indicated.

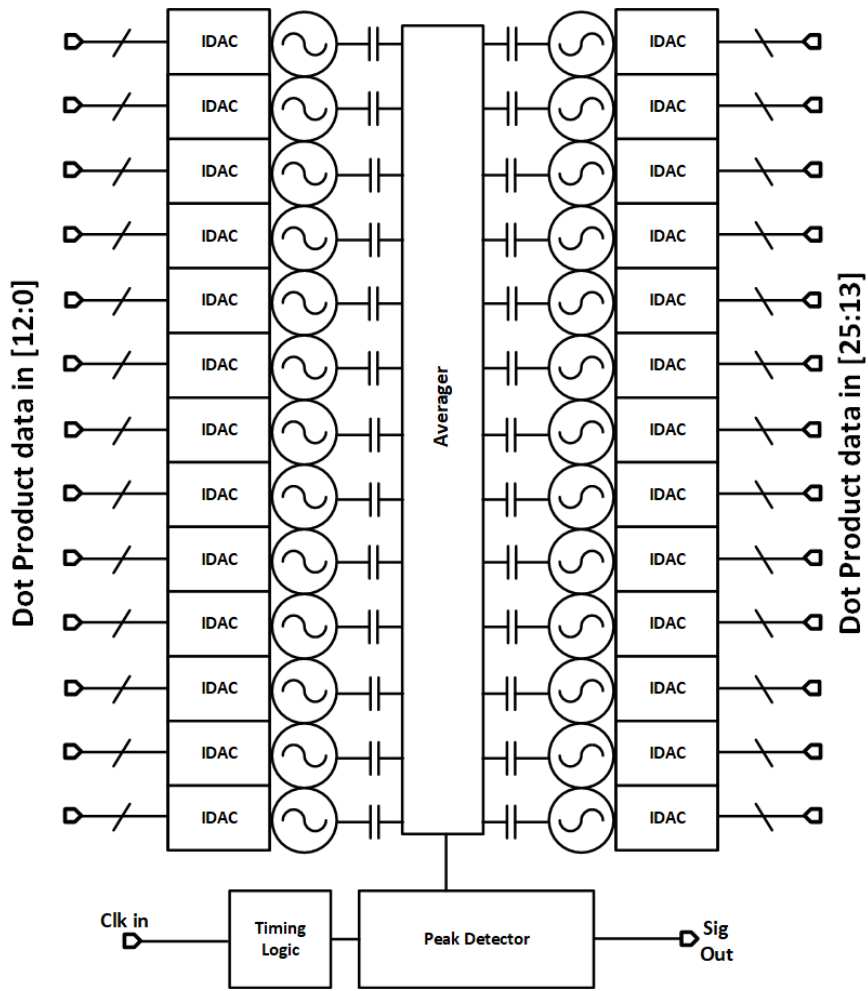

Fig. 2: Scheme of an array of coupled oscillators with current DACs, coupling capacitances, the averager node, and a peak detector.

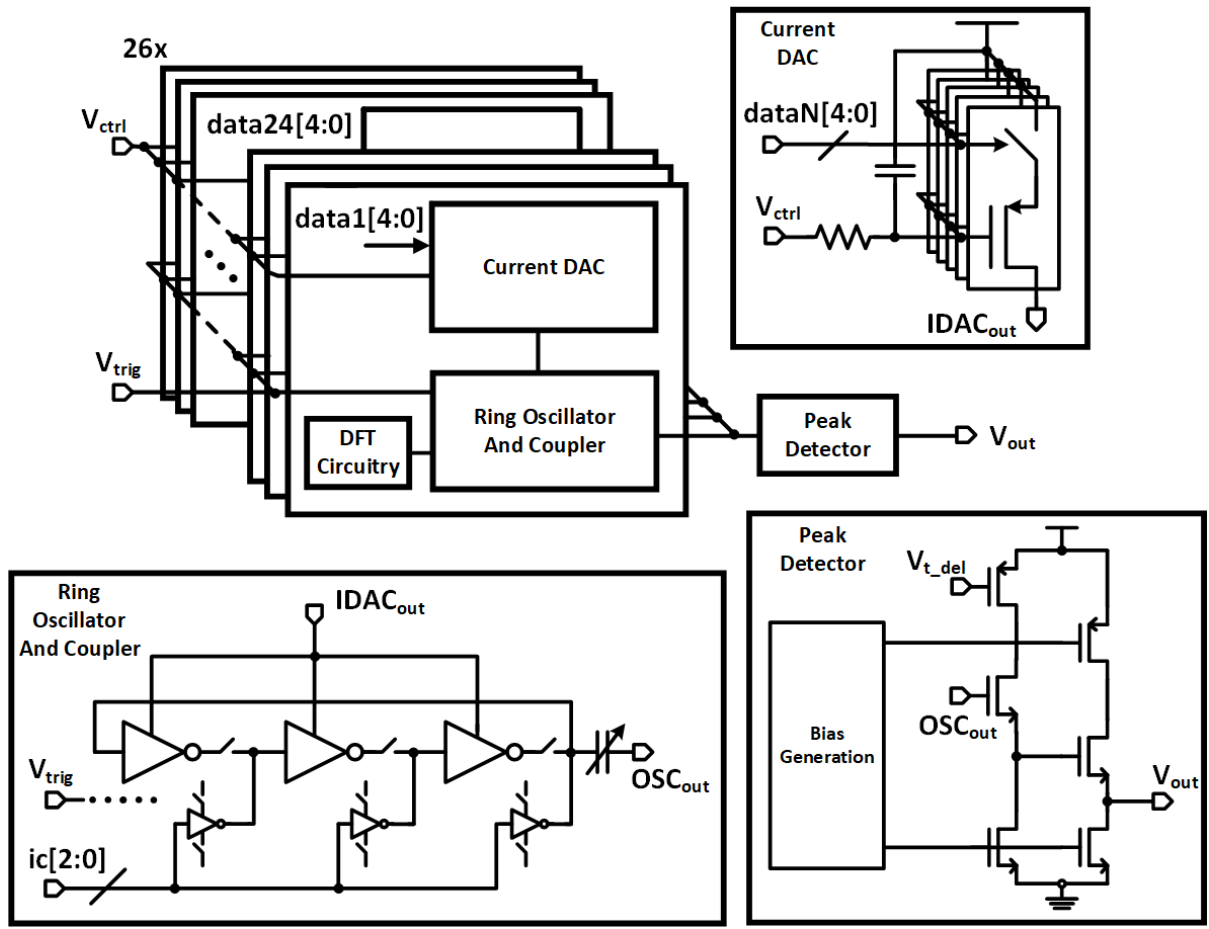

Fig. 3: Circuit diagrams for IDAC, the 3-stage oscillator, the coupling capacitor, and the peak detector.

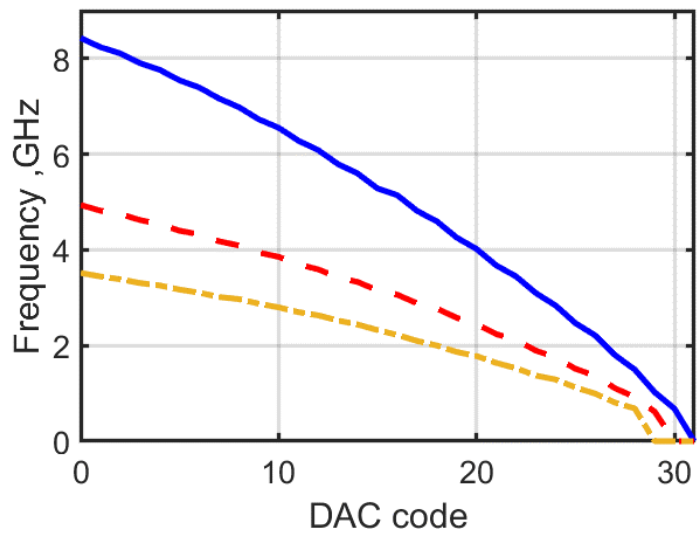

Fig. 4: Measured oscillator frequency vs. DAC code for 3 (blue solid line), 5 (red dashed line), and 7 stages (gold dot-dashed line).

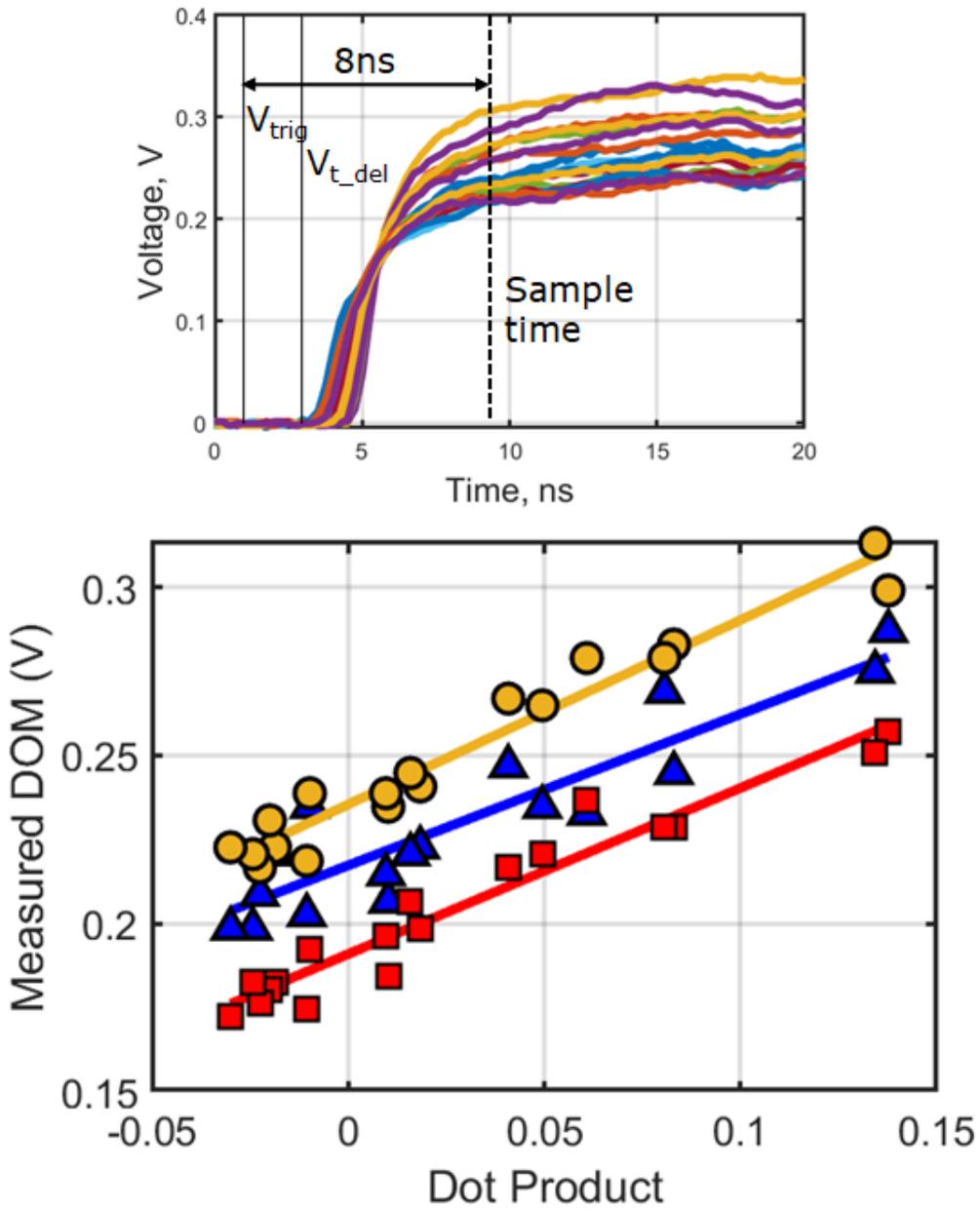

Fig. 5: Measured output of the peak detector for 18 cases of Gabor filters (7stage oscillators) (top). Measured Degree of match data at the sample time for 3 (blue triangles), 5 (red squares), and 7 stages (gold circles) oscillators in the arrays as well as the linear fits to them (bottom).

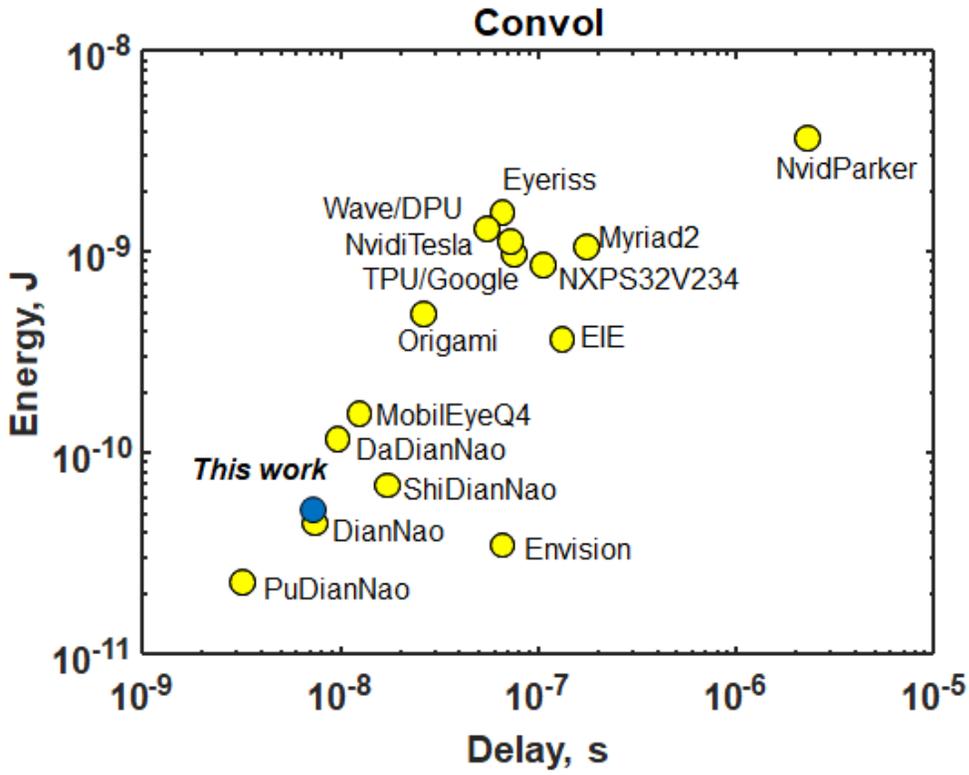

Fig. 6: Benchmarking of inference performance vs. prior neural accelerators.

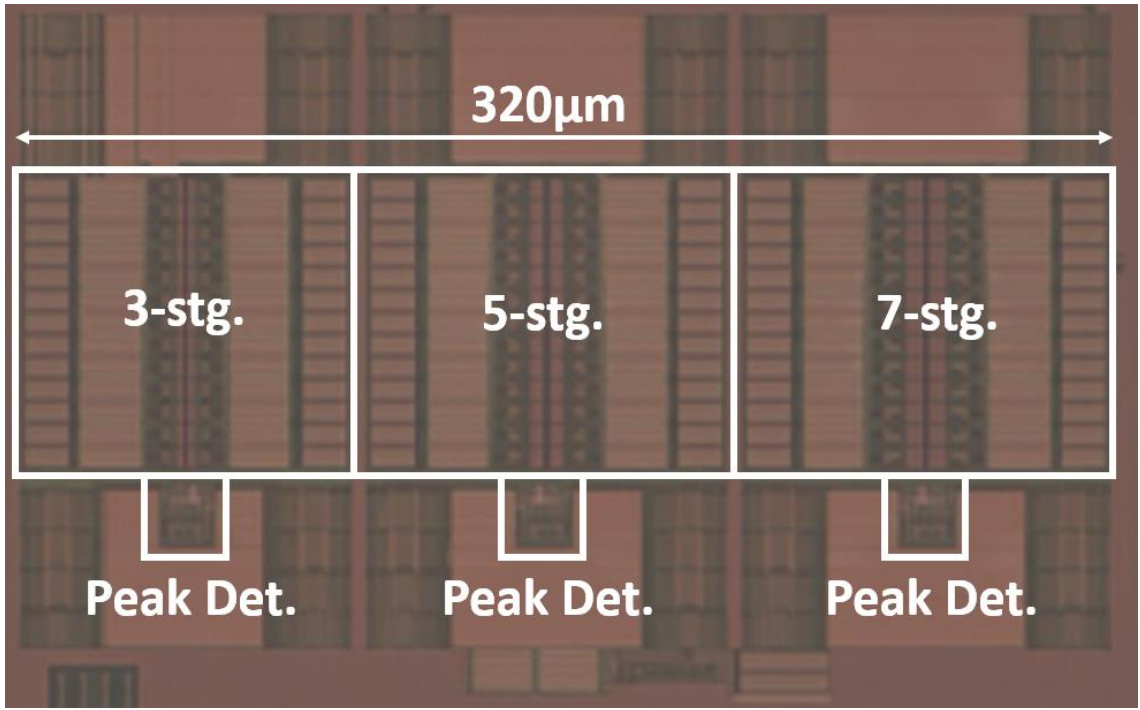

Fig. 7: Die photo of the circuit implementation in 22nm CMOS process showing areas of oscillator arrays.